\documentclass[aps,pra,twocolumn,superscriptaddress,reprint,showpacs]{revtex4-1}  
\usepackage{graphicx}
\usepackage{setspace}
\usepackage{amsmath}
\usepackage{amssymb}
\usepackage{color}
\usepackage{array}
\usepackage{subfigure}
\usepackage{hyperref}
\usepackage{float}
\usepackage{lipsum}

\usepackage[all]{xy}
\newcommand{\RN}[1]{%
  \textup{\uppercase\expandafter{\romannumeral#1}}%
}

\begin{document}
%\begin{CJK*}{GB}{}
\title{Two-qubit controlled-PHASE Rydberg blockade gate protocol via off-resonant modulated driving within a single pulse}

\author{Yuan Sun}
\email[email: ]{sunyuan17@nudt.edu.cn}
\affiliation{Key Laboratory of Quantum Optics and Center of Cold Atom Physics, Shanghai Institute of Optics and Fine Mechanics, Chinese Academy of Sciences, Shanghai 201800, China}
\affiliation{Interdisciplinary Center for Quantum Information, National University of Defense Technology, Changsha 410073, China}
\author{Peng Xu}
\affiliation{State Key Laboratory of Magnetic Resonance and Atomic and Molecular Physics, Wuhan Institute of Physics and Mathematics, Chinese Academy of Sciences -- Wuhan National Laboratory for Optoelectronics, Wuhan 430071, China}
\affiliation{Center for Cold Atom Physics, Chinese Academy of Sciences, Wuhan 430071, China}
\author{Ping-Xing Chen}
\affiliation{Interdisciplinary Center for Quantum Information, National University of Defense Technology, Changsha 410073, China}

\begin{abstract}
Neutral atom array serves as an ideal platform to study the quantum logic gates, where intense efforts have been devoted to improve the two-qubit gate fidelity. We report our recent findings in constructing a different type of two-qubit controlled-PHASE quantum gate protocol with neutral atoms enabled by Rydberg blockade, which aims at both robustness and high-fidelity. It relies upon modulated driving pulse with specially tailored smooth waveform to gain appropriate phase accumulations for quantum gates. The major features include finishing gate operation within a single pulse, not necessarily requiring individual site addressing, not sensitive to the exact value of blockade shift while suppressing population leakage error and rotation error. We anticipate its fidelity to be reasonably high under realistic considerations for errors such as atomic motion, laser power fluctuation, power imbalance, spontaneous emission and so on. Moreover, we hope that such type of protocol may inspire future improvements in quantum gate designs for other categories of qubit platforms and new applications in other areas of quantum optimal control.
\end{abstract}
\pacs{}
\maketitle
%\end{CJK*}

%%%%

\section{Introduction}

Efficient, robust and high-fidelity two-qubit controlled-PHASE gate has become one of the central topics in the research frontier of quantum information with neutral atoms, which is not only important for quantum logic processing \cite{PhysRevLett.104.010503, PhysRevA.92.022336, PhysRevLett.119.160502}, but also crucial for quantum simulation \cite{nature24622} and quantum metrology \cite{nphoton.2011.35, RevModPhys.89.035002}. Rydberg blockade \cite{nphys1178, RevModPhys.82.2313, J.Phys.B.49.202001} emerges as one essential tool for this purpose, where the rapid progress in related studies over the past two decades, both theoretical and experimental, has already found many key advances in quantum information science and technology with neutral atoms \cite{PhysRevA.66.065403, PhysRevLett.99.260501, PhysRevLett.107.093601, PhysRevLett.107.133602, PhysRevLett.109.233602, Dudin887Science, PhysRevLett.110.103001, PhysRevLett.113.053601, PhysRevA.92.022336, PhysRevLett.117.223001, PhysRevLett.119.160502}. One prominent feature of neutral atoms is that they serve as not only good candidates for qubit registers, but also good choices for quantum interface with light, where Rydberg blockade has been deemed as a critical element on both sides \cite{PhysRevLett.109.233602, PhysRevLett.112.040501, PhysRevLett.115.093601, PhysRevLett.119.113601, PhysRevA.91.030301, Hao2015srep, PhysRevA.93.040303, PhysRevA.93.041802, PhysRevA.94.053830, PhysRevA.95.041801, OPTICA.5.001492, ISI:000457492900011}. 

Ever since the seminal paper of Ref. \cite{PhysRevLett.85.2208} which pioneered in the field of quantum computing with neutral atoms, many mechanisms of constructing two-qubit controlled-PHASE gate via Rydberg blockade have been studied extensively so far. Typically, a fast and robust gate mechanism requires relatively strong blockade shift. For those feasible gate protocols readily compatible with the currently mainstream atomic physics experimental platforms, it seems to us that they can be approximately divided into four categories. Category \RN{1}. Rydberg blockade gate with the so-called $\pi$-gap-$\pi$ pulse sequence, which comes from the initial gate designs in Ref. \cite{PhysRevLett.85.2208}. It attracts persistent theoretical interests and serves as the current mainstream blueprint for serious experimental efforts, although it requires individual site addressing. Recent progress has suggested that the gate operation can be performed on the order of several hundred nano-seconds \cite{PhysRevA.88.062337, PhysRevA.92.042710, PhysRevA.94.032306, PhysRevA.96.042306}. Nevertheless, gate fidelities reported from several labs are still at a little distance away from 99\%, which may be partly due to several potential shortcomings embedded in this type of protocol, including a stringent requirement on ground-Rydberg coherence. Category \RN{2}. Rydberg dressing, which was first conceived in the context of quantum gases \cite{PhysRevLett.85.1791}. The blockade effect can also be explored via Rydberg dressing of the ground state atoms \cite{PhysRevA.65.041803, PhysRevA.82.033412}, which can in turn yield a two-qubit controlled-PHASE gate protocol \cite{PhysRevA.91.012337, nphys3487}. However, it usually needs a relatively long gate time which foreshadows gate fidelity due to the finite Rydberg level life time. Besides its role in the universal quantum computing, Rydberg dressing is suitable for implementing adiabatic quantum computation such as quantum annealing \cite{PhysRevA.87.052314}, and finds important applications in constructing multi-qubit quantum simulator \cite{nphys3835, nature24622}. Category \RN{3}. Rydberg anti-blockade gate \cite{PhysRevLett.98.023002, PhysRevLett.111.033607}. Such gate protocols usually requires the exact knowledge of the blockade shift \cite{PhysRevApplied.7.064017}, and are practically more sensitive to fluctuations of the relative motion between two atoms. Category \RN{4}. Protocols with simplified pulse sequence but more theoretical compromises, whose best achievable fidelity is less than ideal but relatively straightforward for experimental demonstration. For example, recently Ref. \cite{EPL-113-4-40001} discussed such a gate protocol with a single square pulse driving ground-Rydberg transition. The major challenge for those protocols is to improve the highest theoretical fidelity limit to fit scalable purpose or fault-tolerant quantum computing. 

Over the years, intense efforts have been devoted to analyzing performances of Rydberg blockade gate \cite{PhysRevA.72.022347, PhysRevA.77.032723, RevModPhys.82.2313, PhysRevA.94.032306, PhysRevA.96.042306}, where both the protocol's inherent physical limitations and technical imperfections have been taken into consideration. Very often, techniques of adiabatic passage \cite{PhysRevA.94.062307, PhysRevA.97.032701}, including STIRAP \cite{PhysRevA.89.030301, PhysRevA.90.033408}, are employed together with the $\pi$-gap-$\pi$ \cite{PhysRevA.96.042306} and Rydberg dressing \cite{PhysRevA.91.012337} gate protocols. Tuning the F\"{o}rster resonance with dc electric fields \cite{nphys3119} or microwave have been also anticipated to facilitate gate performance. Nevertheless, experimental fidelities from those two-qubit gate mechanisms seem relatively less optimistic at this moment, despite the overall rapid progress in this field. Therefore, there exists strong demand for further explorations in gate protocols of potentially different recipes which may overcome known inconveniences in existing protocols.

In this article, we report our recent progress in theoretically devising and analyzing a Rydberg blockade type of two-qubit controlled-PHASE gate protocol for neutral atoms via Rydberg blockade, whose working principles rely upon atom-light interaction with a single off-resonant modulated laser pulse driving the ground-Rydberg transition. The modulation of the pulse waveform is engineered such that within the required fidelity, both the control and target atoms will return to original state no matter the blockade takes place or not, while gaining the correct phases as required by the two-qubit gate. Approximately speaking, this type of protocol combines the advantages of the $\pi$-gap-$\pi$ gate and Rydberg dressing gates in a hybrid form, while avoiding shelving steady population on Rydberg state of control atom for a finite time gap during gate operation and gaining more robustness against noises. The rest of this article is organized as follows, first we present the basic mechanisms of our gate protocol, then we analyze and discuss the results, and finally we conclude the article. Relevant technical details, specifics of derivations and extra examples are included in the supplementary material.

\section{Basic mechanism}

\begin{figure}[t]
\centering
\includegraphics[width=\linewidth]{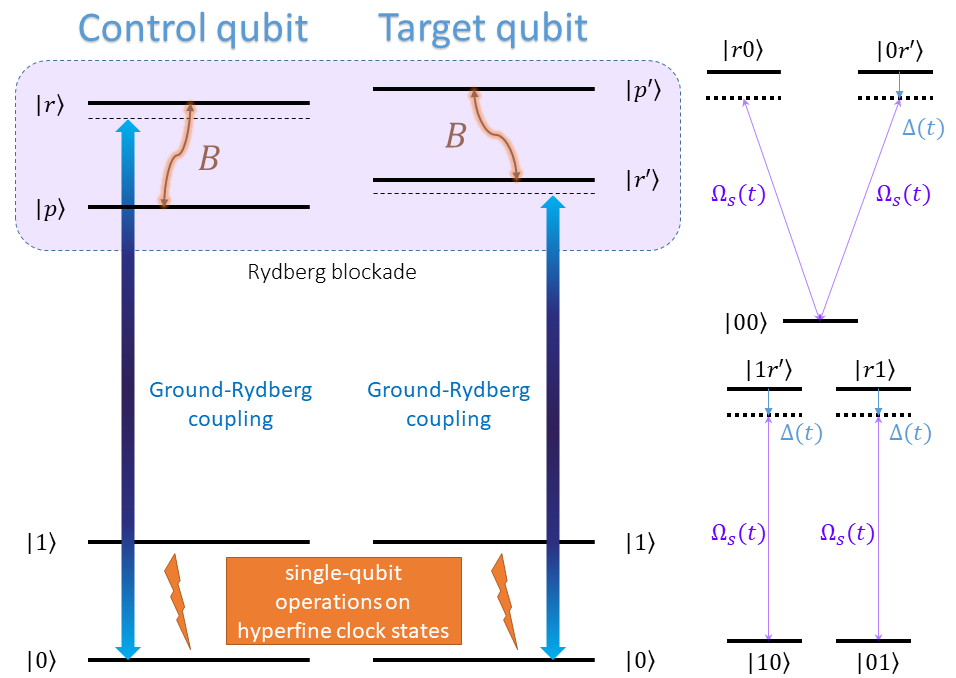}
\caption{Schematic of atomic structure for the Rydberg phase gate under investigation. On the left: the relevant atomic states including the Rydberg blockade between $|r\rangle$ and $|r'\rangle$, where the lasers are driving $|0\rangle \leftrightarrow |r\rangle$ on control atom and $|0\rangle \leftrightarrow |r'\rangle$ on target atom; on the right: under ideal blockade situation, the linkage pattern for states participating the ground-Rydberg transitions $|01\rangle, |10\rangle$ and $|00\rangle$. See Morris-Shore transform at Ref. \cite{PhysRevA.27.906} for a better explanation of comprehending linkage structures. State $|11\rangle$ does not participate the prescribed interactions and stays unchanged through the process. Rydberg states $|r\rangle$ and $|r'\rangle$ may be the same or different, depending on the choice of F\"orster resonance structure. }
\label{fig:basic_0}
\end{figure}

We start with the basic ingredients, where relevant atomic states of the atom-light interaction are shown in Fig. \ref{fig:basic_0}. The qubit basis states of the atoms may be represented by a pair of long-lived hyperfine ground clock states for typical alkali atoms, which can be manipulated by external microwave field or optical stimulated Raman transition \cite{PhysRevLett.114.100503, PhysRevA.92.022336, J.Phys.B.49.202001}. Modulated laser pulses will be applied to drive the ground-Rydberg transitions of the control and target atoms. The consequences of the required operation can be abstracted into two aspects: the boomerang condition that the population returns with unity probability and the antithesis condition that the accumulated phases achieve controlled-Z (C-Z) gate result. When combined with a local Hadamard gate on the target qubit atom ($\pi/2$ rotation for transition $|0\rangle \leftrightarrow |1\rangle$) before and after the controller-PHASE gate, this leads to the universal controlled-NOT gate \cite{PhysRevLett.85.2208, RevModPhys.82.2313, J.Phys.B.49.202001}. If $|r\rangle, |r'\rangle$ are the same state, then individual atom addressing may not be mandatory and the experiment can be operated through one single laser. For simplicity, throughout this article the condition of symmetric driving will be presumed, namely both the qubit atoms will receive the same Rabi frequency and detuning in their effective ground-Rydberg transition couplings. 

Assuming the presence of ideal Rydberg blockade such that double Rydberg excitation $|rr'\rangle$ is impossible. Define the state $|R\rangle = (|r0\rangle+|0r'\rangle)/\sqrt{2}$, there exist three types of couplings: $|10\rangle\leftrightarrow|1r\rangle$, $|01\rangle\leftrightarrow|r1\rangle$ with Rabi frequency $\Omega_s$ and $|00\rangle \leftrightarrow |R\rangle$ with Rabi frequency $\sqrt{2}\Omega_s$, as can be seen in the linkage structure of Fig. \ref{fig:basic_0}. The goal is to find an atom-light interaction process such that the induced changes in the wave functions conforms to the two-qubit phase gate after the interaction is over. More specifically, let $(C_0, C_r)$ denote the wave function for the ground-Rydberg transition of $|10\rangle$ or $|01\rangle$, and let $(X_0, X_R)$ denote the wave function for the ground-Rydberg transition of $|00\rangle$. The problem reduces to determine proper and feasible $\Omega_s, \Delta$ for the time evolution:
\begin{subequations}
\label{ideal_blockade_eom}
\begin{align}
i\frac{d}{dt} \begin{bmatrix}C_0\\C_r\end{bmatrix}
=
\begin{bmatrix}
0 & \frac{1}{2}\Omega_s  \\
\frac{1}{2}\Omega_s^*  & \Delta 
\end{bmatrix}
\cdot
\begin{bmatrix}C_0\\C_r\end{bmatrix};\\
i\frac{d}{dt} \begin{bmatrix}X_0\\X_R\end{bmatrix}
=
\begin{bmatrix}
0 & \frac{\sqrt{2}}{2}\Omega_s  \\
\frac{\sqrt{2}}{2}\Omega_s^*  & \Delta 
\end{bmatrix}
\cdot
\begin{bmatrix}X_0\\X_R\end{bmatrix}.
\end{align}
\end{subequations}

It turns out, appropriate solutions may be obtained, where the practical task becomes to find them out and examine their properties. Our tactics involve careful refining efforts for modulations from heuristic approaches. More specifically, first we design waveforms under assumption of perfect adiabatic time evolution process in Eq. \eqref{ideal_blockade_eom}, and then perform optimizations to suppress the non-adibatcity effects \cite{SuppInfo}, where numerical tools serves an essential role in this process.

Except for technical noises, we think that two main types of intrinsic errors exist: the population leakage error due to spontaneous emission of Rydberg levels during interaction, and the rotation error due to the less than ideal Rydberg blockade with double Rydberg excitation. Nevertheless, with properly tailored smooth pulses, the mechanism of adiabatically tracking two-atom dark state gets implicitly triggered under the presence of dipole-dipole exchange interaction $|rr'\rangle \leftrightarrow |pp'\rangle$ \cite{SuppInfo}. Therefore, the rotation error will be suppressed as we may observe in later discussions.

\begin{figure}[b]
\centering
% from notebook B
\fbox{\includegraphics[width=0.95\linewidth]{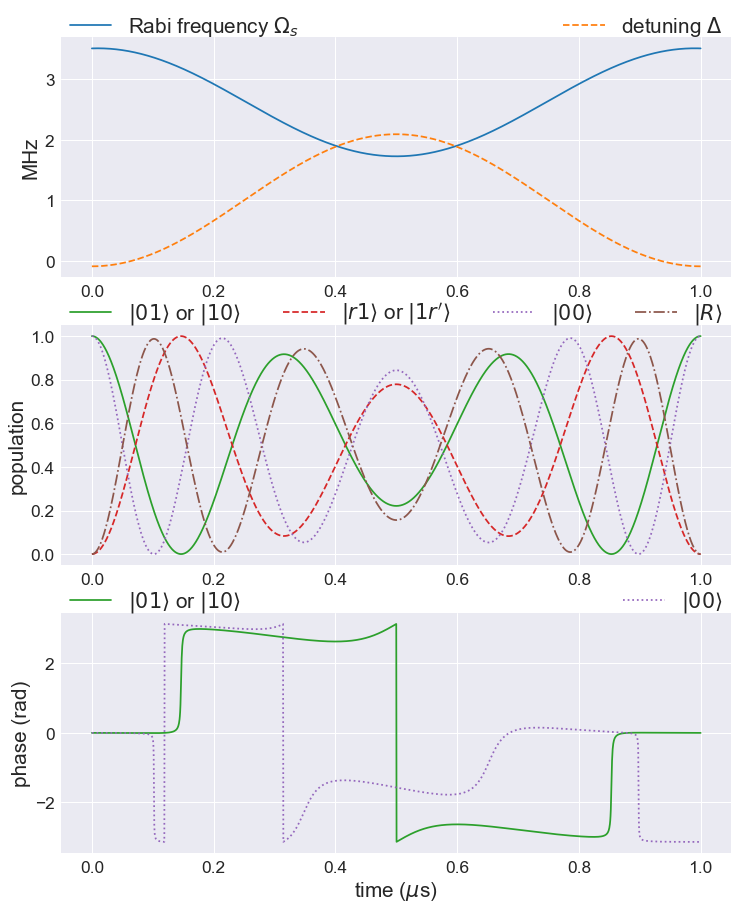}}
\caption{Numerical simulation of the time evolution. Modulation is configured as stated in the text, while $B=2\pi \times 500$ MHz, $\delta_p = 2\pi \times -3$ MHz. The first graph shows the waveform, the second graph shows the population on different atomic states, while the last graph shows the phase accumulation of the atomic wave function during the process. The purpose is C-Z gate, and the fidelity of this example is 0.99997. To evaluate fidelity, we first numerically calculate the outcome wave functions from the time evolution. Then, with respect to the four basis states $|00\rangle, |01\rangle, |10\rangle, |11\rangle$, we acquire the 4 by 4 transform matrix $U$ representing our gate operation. Then the fidelity may be calculated as $F = ( \text{Tr}(MM^\dagger) + |\text{Tr}(M)|^2 )/20$, with $M = U_\text{C-Z}^\dagger U$ with $U_\text{C-Z}$ being the transform matrix of an ideal C-Z gate.}
\label{fig:hybrid_modulation_0}
\end{figure}

\section{Results and discussions}

For $|01\rangle$ and $|10\rangle$, the dynamics amounts to nothing more than a two-level system made from ground-Rydberg transition with time-dependent Rabi frequency $\Omega_s(t)$ and detuning $\Delta(t)$. On the other hand, for $|00\rangle$, its dynamics actually probes the Rydberg dipole-dipole interaction, whose linkage pattern may be summarized as $|00\rangle \leftrightarrow |R\rangle \leftrightarrow |rr'\rangle \leftrightarrow |pp'\rangle$. In order to quantitatively describe the F\"oster resonance structure of $|rr'\rangle \leftrightarrow |pp'\rangle$, we assume that the coupling strength is $B$ and the small F\"{o}rster energy penalty term is $\delta_p$ for $|pp'\rangle$. The interaction Hamiltonian is then:
\begin{eqnarray}
\label{eq:Rydberg_Hamiltonian}
H_I/\hbar = &\frac{\sqrt{2}}{2}\Omega_s|R \rangle \langle 00| + \frac{\sqrt{2}}{2}\Omega_s|rr'\rangle \langle R| + \text{H.c.}
\nonumber\\
&+ \Delta |R\rangle\langle R| + 2\Delta |rr'\rangle\langle rr'|, 
\end{eqnarray}
where we have already included rotating wave approximation. We include Rydberg blockade as: 
\begin{equation}
\label{eq:blockade_Hamiltonian}
H_F/\hbar = B|pp'\rangle \langle rr'\rangle + \text{H.c.} 
+ \delta_p |pp'\rangle\langle pp'|.
\end{equation}

Following the prescribed recipes, we have obtained two categories of waveforms that yield two-qubit phase gate with satisfying performances. One of them requires amplitude and frequency modulations simultaneously, while the other one only requires amplitude modulation with a constant detuning.

For the waveform of both amplitude and frequency modulations, we make the designing goal a little more strict than necessary, such that the aim is for a standard C-Z gate where the population returns with a phase change of 0 for $|01\rangle$ and $|10\rangle$ and a phase change of $\pi$ for $|00\rangle$ after interaction. Beginning with heuristic approaches \cite{SuppInfo}, we find out the sinusoidal modulations fit well for this category after the refining work on the time evolution details. In particular, we have selected a set of waveforms described in the following:
\begin{subequations}
\label{eq:waveform_v7e}
\begin{align}
\Omega_s (t) = \Omega_0 + \Omega_1 \cos(2\pi t/T_g) + \Omega_2 \sin(\pi t/T_g); \\
\Delta (t) = \Delta_0 + \Delta_1 \cos(2\pi t/T_g) + \Delta_2 \sin(\pi t/T_g).
\end{align}
\end{subequations}

Via optimizations under the prescribed constraints, we have identified a set of values as: $\Omega_0=2.564, \Omega_1=0.950, \Omega_2=0.116, \Delta_0=1.004, \Delta_1=-1.093, \Delta_2=-0.002$; all coefficient units are MHz, and the gate time $T_g$ is set as 1 $\mu$s. To demonstrate the detailed dynamics of the system with respect to the Hamiltonian of Eq. \eqref{eq:Rydberg_Hamiltonian}, we present the numerical simulation results in Fig. \ref{fig:hybrid_modulation_0} without considerations for spontaneous emissions of Rydberg levels and technical noises. The modulation does not involve unreasonable high frequency components, and the atomic wave function does not go through `sudden' change during the course of gate operation. We have intentionally chosen a symmetric waveform, which simplifies the calculation process but is not mandatory. For dynamics associated with $|00\rangle$, although the situation is more complicated than the two-level system considered in Eq. \eqref{ideal_blockade_eom}, a clear signature is that the population returns almost ideally with no significant portion trapped in the Rydberg level, thanks to the adiabatic dark state driving mechanism in Rydberg blockade effect \cite{SuppInfo}.

For the other category of only amplitude modulation, it is preferred that the pulse starts and ends at zero intensity. Among several candidate waveforms, we are particularly interested in the ones of relatively less complexities, such as:
\begin{subequations}
\label{eq:waveform_v22a}
\begin{align}
&\Omega_s (t) = \sum_{\nu=1}^{4} \beta_\nu \big(b_{\nu, n}(t/T_g) + b_{n-\nu, n}(t/T_g) \big); \\
&\Delta (t) = \Delta_0 \equiv \textit{ constant};
\end{align}
\end{subequations}
where $b_{\nu, n}$ is the $\nu$th Bernstein basis polynomials of degree $n$ \cite{SuppInfo}, we set $n=8$ and we again intentionally configure a symmetric waveform. The result we pursue is in fact a controlled-PHASE gate, and local sing-qubit phase rotation is required if we want conversion into C-Z gate. The associated phase constraint is:
\begin{equation}
\label{C-PHASE_constraint}
\phi_{11} = \pm \pi - \phi_{00} + \phi_{01} + \phi_{10},
\end{equation}
where $\phi_{01} + \phi_{10} - \phi_{00} = \pm\pi$ if $\phi_{11} = 0$.

\begin{figure}[t]
\centering
% from notebook B2a
\fbox{\includegraphics[width=0.95\linewidth]{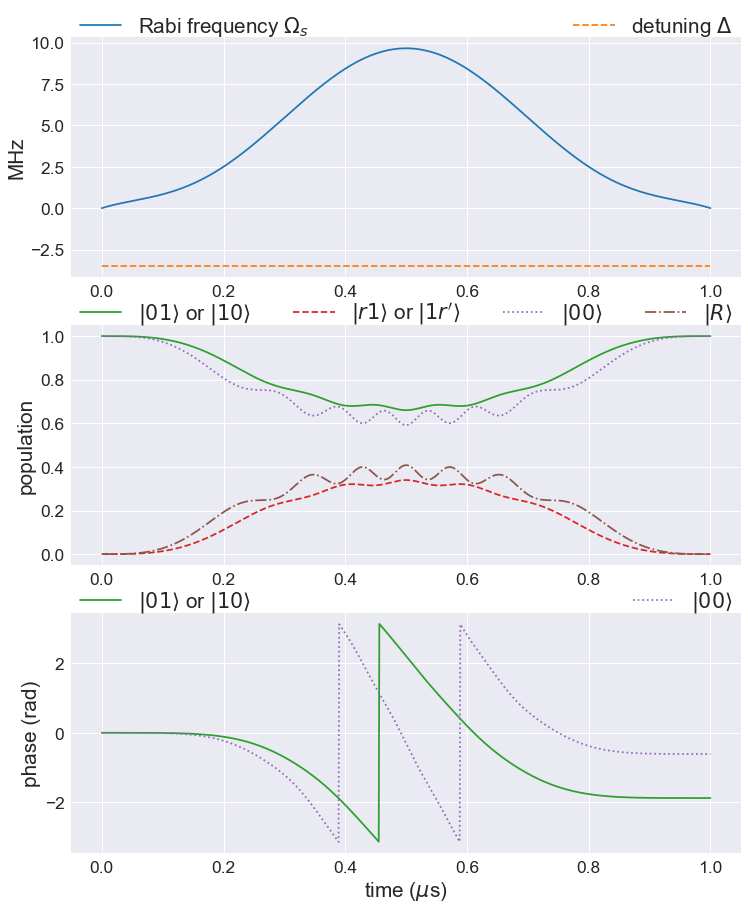}}
\caption{Numerical simulation of the time evolution with amplitude modulated pulse. The waveform is set as Eq. \eqref{eq:waveform_v22a}, while $B=2\pi \times 500$ MHz, $\delta_p = 2\pi \times -3$ MHz. The first graph shows the waveform, the second graph shows the population on different atomic states, while the last graph shows the phase accumulation of the atomic wave function during the process. After appropriate local phase rotations to adjust it to a standard C-Z gate, its gate error $\mathcal{E}$ is way below $1\times10^{-5}$, defined as $\mathcal{E} = 1- F$. Spontaneous emissions of Rydberg levels and technical noises are not considered here.}
\label{fig:amplitude_modulation_0}
\end{figure}

Next, we seek a set of values leading to appropriate phase gate performance. After optimization efforts, for gate time $T_g$ is set as 1 $\mu$s, we have reached a set of satisfying parameters, $\beta_1=1.419, \beta_2=0, \beta_3=5.076, \beta_4=13.425, \Delta_0=-3.512$; all coefficient units are MHz. The corresponding numerical simulation is shown in Fig. \ref{fig:amplitude_modulation_0}. The singly-excited Rydberg state $|R\rangle$ is not heavily populated throughout the interaction process, which does not share the same behavior as the obvious feature of quantum Rabi oscillation in Fig. \ref{fig:hybrid_modulation_0}. This is due to the difference in the underlying physics mechanisms between those two cases, where Fig. \ref{fig:hybrid_modulation_0} shares similarities with a typical quantum Rabi oscillation and Fig. \ref{fig:amplitude_modulation_0} shares similarities with adiabatic rapid passage, and this may be observed from their paths on Bloch sphere. Nevertheless, both approaches are suitable choices for the purpose of two-qubit phase gate.

We observe that the mechanism of adiabatically tracking the two-atom dark state also plays an essential role here \cite{SuppInfo}. The amplitude modulation does not only introduce the correct change in atomic wave functions for a phase gate with respect to Eq. \eqref{ideal_blockade_eom} and Eq. \eqref{C-PHASE_constraint}, but also helps to suppress the rotational error. In other words, major limitations on the attainable fidelity are anticipated to mostly come from spontaneous emissions, modulation imperfections and technical noises.

\begin{figure}[t]
\centering
\includegraphics[width=0.95\linewidth]{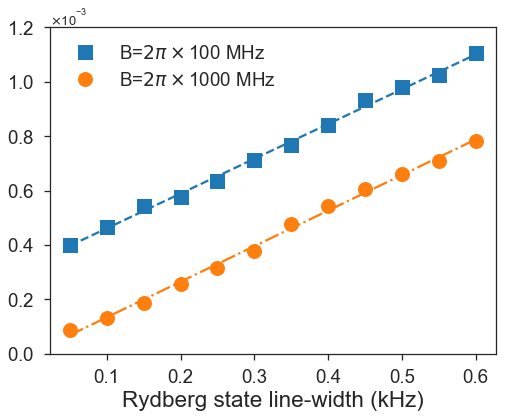}
\caption{Numerical simulation for gate error $\mathcal{E}$ with gate time set as 1 $\mu$s and the waveform set as Fig. \ref{fig:amplitude_modulation_0}. Each data point is extracted from 200,000 MCWF trajectories. Two values of $B$ are considered, all with $\delta_p = 2\pi \times -3$ MHz. For simplicity, the spontaneous decay rates of all Rydberg states are taken as the same. Fittings to a straight lines indicates that the linear relation holds well between Rydberg decay rate and gate error, for the parameter range we are interested in.}
\label{fig:fidelity_mcwf_0}
\end{figure}

Furthermore, we need to estimate the influences of the major intrinsic error source of spontaneous emission. We carry out numerical evaluations by resorting to quantum jump approach \cite{PhysRevLett.68.580, RevModPhys.70.101}, also known as Monte-Carlo wave function (MCWF). The result is shown in Fig. \ref{fig:fidelity_mcwf_0}, where we compute gate error as a function of the Rydberg decay rates. We deduce that in principle the Rydberg levels' spontaneous emission is the dominating theoretical limiting factor in achieving high-fidelity, provided the Rydberg blockade shift is strong enough.  

We have also investigated the influence to gate performance from realistic imperfections commonly encountered in experiments, including amplitude fluctuations in the laser pulse, residual thermal motion of the cold atoms and laser power imbalance at the two qubit sites \cite{SuppInfo}. We observe that the gate protocol is robust against these types of disturbance as long as they are kept at a reasonably low level. When considering the off-resonant driving for the ground-Rydberg transition, we've concentrated on the few states that are directly involved in the mechanism. Realistically, the situation is more complicated due to the atom's many other levels, which may introduce various sources of extra ac Stark shifts and decoherences \cite{PhysRevA.72.022347, PhysRevA.94.032306}.

Several major characteristics are worth mentioning here. It may work without individual addressing on the qubit atoms. No microwave is required to drive the Rydberg-Rydberg transitions and henceforth it saves trouble of the complicated microwave electronic equipment and antennas. It does not require the exact knowledge of the magnitude of the Rydberg blockade shift. Its working condition does not involve far off-resonance detuning as that of Rydberg dressing and therefore the gate may be designed to fast operation below 1 $\mu$s with respect to realistic experimental apparatus parameters. Carrying out two-qubit entangling gate within a single continuous shaped driving pulse has already become a common practice in other platforms such as the superconducting qubit \cite{PhysRevLett.107.080502}, and we think it is beneficial to design a counterpart for neutral atom platform.

\section{Conclusion and outlook}

We have systematically presented our recent results in designing two-qubit controlled-PHASE Rydberg blockade gate protocol for neutral atoms via off-resonant modulated driving within a single pulse. In principle, the same guidelines may be extended to help construct a generic three-qubit gate such as Toffoli gate. 

On the other hand, we believe that the result is not unique and a full characterization of accessible solutions remains an open problem. We are also looking forward to a few other future refinements, including the search for a faster gate operation, further suppression of population leakage, stronger robustness against environmental noises, and more user-friendly parameter setting. Error correction mechanism \cite{PhysRevLett.117.130503} for our gate protocol is also part of the long term goal.

For applications with readily available hardware in immediate future \cite{PhysRevA.92.022336, PhysRevLett.119.160502},  we expect that $\gtrsim 99$\% fidelity may be obtained for two-qubit gate in 1D, 2D or 3D atomic arrays with a gate time less than 1 $\mu$s. Our faith with the Rydberg blockade gate is that high-fidelity ground-Rydberg Rabi oscillation shall be directly translated into high-fidelity controlled-PHASE gate. We also anticipate that our work will help the endeavors for the ensemble qubit approach \cite{PhysRevLett.115.093601, PhysRevLett.119.180504} and the Rydberg-mediated atom-photon controlled-PHASE gate \cite{Hao2015srep, PhysRevA.93.040303, PhysRevA.94.053830, OPTICA.5.001492}. 

\begin{acknowledgements}

The authors gratefully acknowledge the funding support from the National Key R\&D Program of China (under contract Grant No. 2016YFA0301504 and No. 2016YFA0302800). The authors also acknowledge the hospitality of Key Laboratory of Quantum Optics and Center of Cold Atom Physics, Shanghai Institute of Optics and Fine Mechanics. The authors gratefully thank the help from Professor Liang Liu, Professor Mingsheng Zhan  and Professor Mark Saffman who essentially make this work possible. The authors also thank Professor Xiaodong He, Professor Dongsheng Ding and Professor Tian Xia for enlightening discussions.

\end{acknowledgements}

\bibliographystyle{apsrev4-1}

\renewcommand{\baselinestretch}{1}
\normalsize

%\clearpage%
%\phantomsection%
%\addcontentsline{toc}{chapter}{\numberline{}{Bibliography}}%
\bibliography{duet_ref}

\end{document}